\documentclass[a4paper]{jpconf}
\usepackage{graphicx}
\begin{document}
\title{Exploring the spectrum of QCD using the lattice}

\author{\textit{Hadron Spectrum Collaboration}: John Bulava$^1$, Saul Cohen$^2$, Jozef Dudek$^2$, Robert
  Edwards$^2$, Eric Engelson$^3$, Justin Foley$^1$, Balint Joo$^2$,
  Jimmy Juge$^4$, Huey-Wen Lin$^2$, Nilmani Mathur$^5$, Colin
  Morningstar$^1$, Mike Peardon$^6$, $^\dagger$David Richards$^2$, Sinead
  Ryan$^6$, Christopher Thomas$^2$, Anthony Thomas$^2$, Stephen
  Wallace$^3$}

\address{$^1$~Department of Physics, Carnegie Mellon University, Pittsburgh, PA
  15213, USA}

\address{$^2~$Thomas Jefferson National Accelerator Facility, Newport News,
  VA 23606, USA}

\address{$^3~$Department of Physics, University of Maryland, College Park,
  MD 20742, USA}

\address{$^4~$Department of Physics, University of the Pacific, Stockton,
  CA 95211, USA}

\address{$^5~$Department of Theoretical Physics, Tata Institute of Fundamental Research, Mumbai 400005, India}

\address{$^6~$School of Mathematics, Trinity College, Dublin 2, Ireland}

\ead{$^\dagger$dgr@jlab.org}
\begin{abstract}
The calculation of the spectrum of QCD is key to an
understanding of the strong interactions, and vital if we are to
capitalize on the experimental study of the spectrum.  In this paper,
we describe progress towards understanding the spectrum of
resonances of both mesons and baryons from lattice QCD, focusing in
particular on the resonances of the $I=1/2$ nucleon states, and of
charmonium mesons composed of the heavy charmed quarks.
\end{abstract}
\section{Introduction}
The strong interaction, one of the forces of the Standard Model of
particle and nuclear physics, is responsible for binding both protons
and neutrons into nuclei, and also the primordial gluons and the
lightest quarks into pions, protons, neutrons and other so-called
œôòühadronsœôòý. The interaction arises from a quantum field theory known as
Quantum Chromodynamics (QCD).

In order to really understand QCD and hence test whether it is the
complete theory of the strong interaction, we must know the spectrum
of mesons and baryons that it implies and test those spectra against
high quality data. The complete combined analysis of available
experimental data on the photoproduction of nucleon resonances is the
2009 milestone in Hadronic Physics (HP), and the measurement of the
electromagnetic properties of the low-lying baryons is an HP 2012
milestone.  The observed spectrum of QCD provides little direct
evidence of the presence of the gluons.  However, QCD admits the
possibility of exotic mesonic states of matter in which the gluonic
degrees of freedom are explicitly exhibited.  The search for such
states will be an important component of the
upgraded JLab@12GeV.

Given the intense experimental efforts in hadron spectroscopy, the
need to predict and understand the hadron spectrum from first
principles calculations in QCD is clear.  Hence, an inportant goal of
the effort of the USQCD Collaboration is the study of the
resonance spectrum of QCD.  The remainder of this paper is laid out as
follows.  In the next section, we introduce lattice QCD, and outline
the methodology for looking at resonances on the lattice.  We then
describe recent progress at understanding the resonance spectrum,
focusing first on nucleon resonances and then on the meson sector.  We
conclude with prospects for future calculations.

\section{Lattice QCD}
The interactions of the quarks and gluons at very short distances,
such as those probed at the LHC, are weak, a property known as
œôòüasymptotic freedomœôòý, for which Gross, Politzer, and Wilczek won the
2004 Nobel Prize.  This enables the interactions to be expanded in
terms of a small coupling constant. At longer distances, typical of
the binding of the quarks and gluons into hadrons, the coupling
becomes strong, and the theory highly non-linear.  Here, the only
means of solving, as opposed to modelling, QCD is through numerical
calculations on the lattice.

Lattice gauge calculations solve QCD on a four-dimensional lattice, or
grid, of points in Euclidean space.  The quarks reside on the points
of the grid, whilst the gluons are associated with the links joining
those points.  Lattice calculations proceed through a Monte Carlo
method, in which ensembles of gauge configurations are generated with
a probability distribution prescribed by the Euclidean QCD action.
Lattice QCD calculations have always been at the leading edge of
exploiting the most powerful supercomputing resources available,
helped by the highly regular nature of the problem.

\begin{figure}
\begin{minipage}[b]{0.48\linewidth}\centering
\includegraphics[width=7.5cm]{hadron_spec.eps}\hfill
\vspace{1cm}
\caption{A comparison between the calculated values of the low-lying
  hadron masses and their physical values, obtained in a calculation
  with two flavors of light ($u/d$) quarks and a strange
  quark\protect\cite{Lin:2008pr}. The black and colored bands indicate
  the experimental and lattice values, respectively; the
  width of the band indicates the uncertainty.\label{fig:anisospec}}
\end{minipage}
\hfill
\begin{minipage}[b]{0.48\linewidth}\centering
\includegraphics[width=6.2cm]{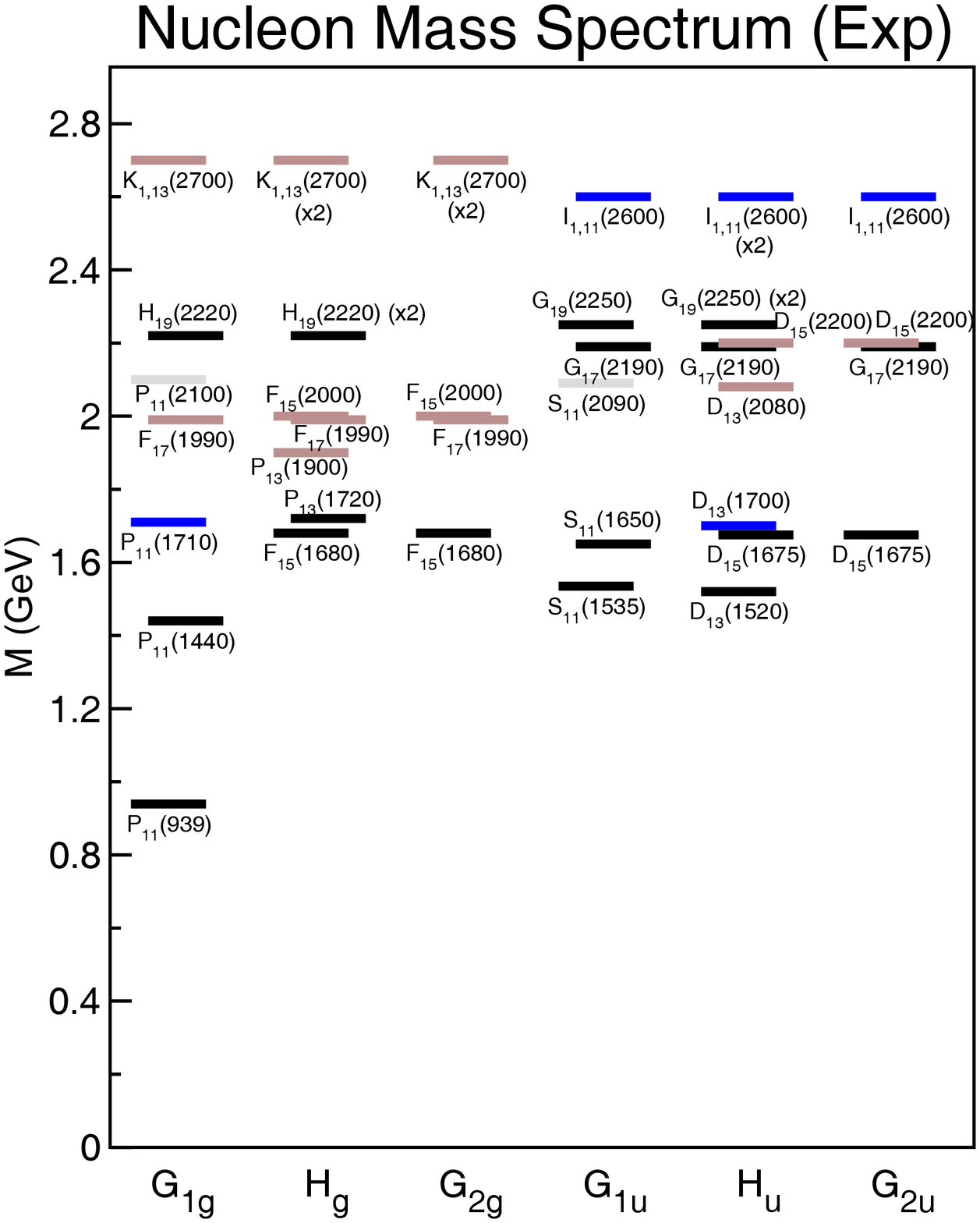}\hfill
\caption{The figure shows the experimentally determined nucleon
  spectrum classified according to the transformation properties under the
  cubic group of the lattice; the lowest-lying state in the $G_{1g}$
  channel is the proton\cite{Lichtl:2006dt}.
\label{fig:exp}}
\end{minipage}
\end{figure}

The calculation of the ground-state spectrum has been a benchmark
calculation of lattice QCD since its inception.
Figure~\ref{fig:anisospec} shows a summary of the low-lying
light-hadron masses compared with their experimental values, measured
on anisotropic clover lattices designed for the study of resonance
spectroscopy\cite{Lin:2008pr}; details of the computational
methodology are given in the poster of Balint Joo.

A comprehensive picture of resonances requires that we go beyond a
knowledge of the ground state mass in each channel, and obtain the
masses of the lowest few states of a given quantum number.  This we
can accomplish through the use of the variational
method\cite{Michael:1985ne,Luscher:1990ck}; we calculate
a matrix of correlator functions
\[
C_{ij}(t) = \sum_{\vec{x}} \langle  O_i(\vec{x}, t) O^{\dagger}_j
(\vec{0}, 0) \rangle \label{eq:corrs_cons},
\]
where $\{  O_i; i = 1,\dots,N \}$ are a basis of interpolating
operators with given quantum numbers.  We then solve the generalized
eigenvalue equation
\[
C(t) u = \lambda(t, t_0) C(t_0) u
\]
to obtain a set of real (ordered) eigenvalues $\lambda_n(t,t_0)$,
where $\lambda_0 \ge \lambda_1 \ge\dots \ge \lambda_{N-1}$.  At large Euclidean times,
these eigenvalues then delineate between the different masses
\[
\lambda_n (t,t_0) \longrightarrow e^{ -M_n (t-t_0)} + O (e^{ - \Delta M_n (t-t_0)}),
\]
where $\Delta M_n = \mbox{min} \{ \mid M_n - M_i \mid : i \ne n \}$.
The eigenvectors $u$ are orthogonal with metric $C(t_0)$, and a
knowledge of the eigenvectors can yield information about the partonic
structure of the states.  Crucial to the application of the
variational method is the use of a basis of interpolating operators that
have a good overlap with the low-lying states of interest. The cubic
lattice employed in our calculations does not admit the full
rotational symmetry of the continuum, but rather the more restricted
symmetry of the octahedral group.  Thus states at rest are classified
according to the irreducible representation (irreps) of the cubic
group, and for spectroscopy calculations, interpolating operators must
be constructed that transform irreducibly under the cubic group;
this task has been the prerequisite for our study both of
baryons\cite{Basak:2005ir,Basak:2005aq}, and of mesons.

\section{Nucleon Resonance Spectrum}
Baryons, containing three quarks, are emblematic of the non-Abelian
nature of QCD, and of the three colors of the theory.  An important
goal in exploring baryons is attempting to discern the effective
degrees of freedom that describe the spectrum; the search for
so-called ``missing resonances'' focuses on whether the spectrum can
be well described by a quark model, or whether an effective theory
with fewer degrees of freedom, such as a quark-diquark picture,
provides a more faithful description of the baryon spectrum.

There are three double-valued irreducible representations of the cubic
group, denoted $G_{1u/g}(2)$, $H_{u/g}(4)$ and $G_{2 u/g}(2)$, where
$g$ and $u$ refer to positive and negative parity, respectively, and
the brackets contain the dimension of the irrep.\; $G_1$ contains
continuum spins $1/2, 7/2,\dots$, $H_g$ spins $3/2, 5/2,\dots$ and
$G_2$ spins $5/2, 7/2, \dots$.  Thus, at any fixed lattice spacing
$a$, a state corresponding to spin-$5/2$ has four degrees of freedom
in $H$, and two in $G_2$, with degeneracies between the energies in
the two irreps emerging in the continuum limit.
Figure~\ref{fig:exp} shows the experimental nucleon spectrum as seen
on the lattice; the limited number of irreps requires that for each
channel we be able to isolate as many energy levels as possible.

\begin{figure}
\begin{center}
\includegraphics[width=5.9cm]{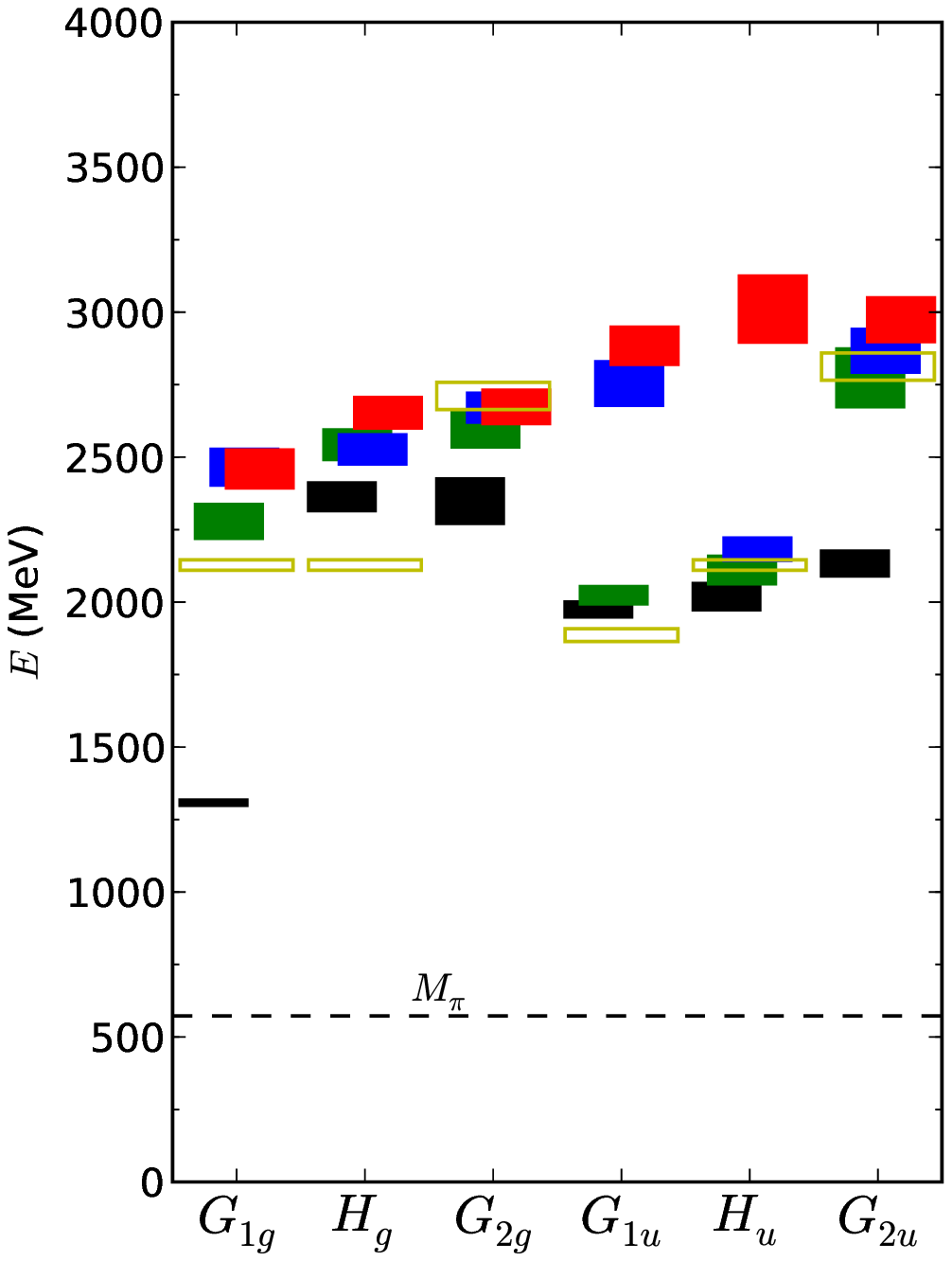}\hfill
\includegraphics[width=5.9cm]{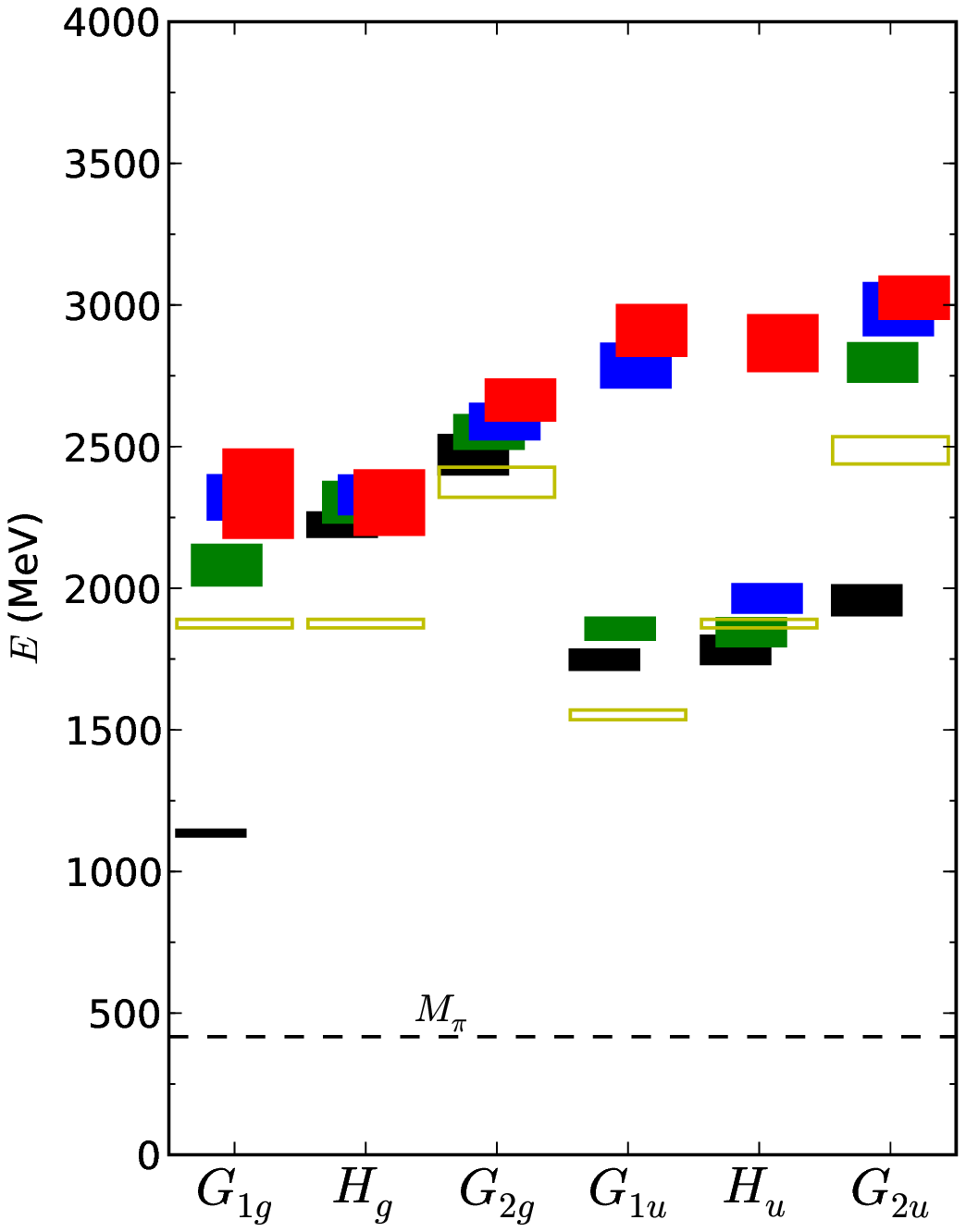}
\caption{The left- and right-hand panels show the spectrum of $I=1/2$
  baryon resonance, indicated by the solid boxes, obtained on $N_f=2$
  Wilson fermion lattices at $m_\pi = 578$ and $416~{\rm MeV}$
  respectively\protect\cite{Bulava:2009jb}; the errors are indicated
  by the vertical width of the box.  The open boxes show the expected
  thresholds for multiparticle states.\label{fig:boxbaryon}}
\end{center}
\vspace{-0.25in}
\end{figure}

The nucleon spectrum has been analysed in a calculation with two
flavors of light Wilson fermions, at two values of the pion
mass\cite{Bulava:2009jb}, building on an earlier calculation in the
quenched approximation to QCD\cite{Basak:2007kj}.  The data are shown
in Figure~\ref{fig:boxbaryon}.  For the first time in a lattice
calculation, we can identify a spin-$5/2$ state, but the multi-hadron
states that should be seen in the spectrum appear elusive; the
calculation of correlation functions that are expected to be sensitive
to these multi-hadron contributions is an important goal for the
collaboration.

\section{Meson Resonances}
The new Hall D of the JLab 12GeV upgrade centers on the study of meson
states produced in photoproduction reactions in the GlueX detector.
Photoproduction has been proposed, within QCD-motivated models, as a
favorable method for the production of ``exotic'' hybrid mesons, those
mesons having $J^{PC}$ outside the set allowed to a
fermion-antifermion pair. The hybrid hypothesis is that an excited
gluonic field in addition to a quark-antiquark pair can give rise to
these quantum numbers.  

\begin{figure}
\begin{center}
\includegraphics[width=7.5cm]{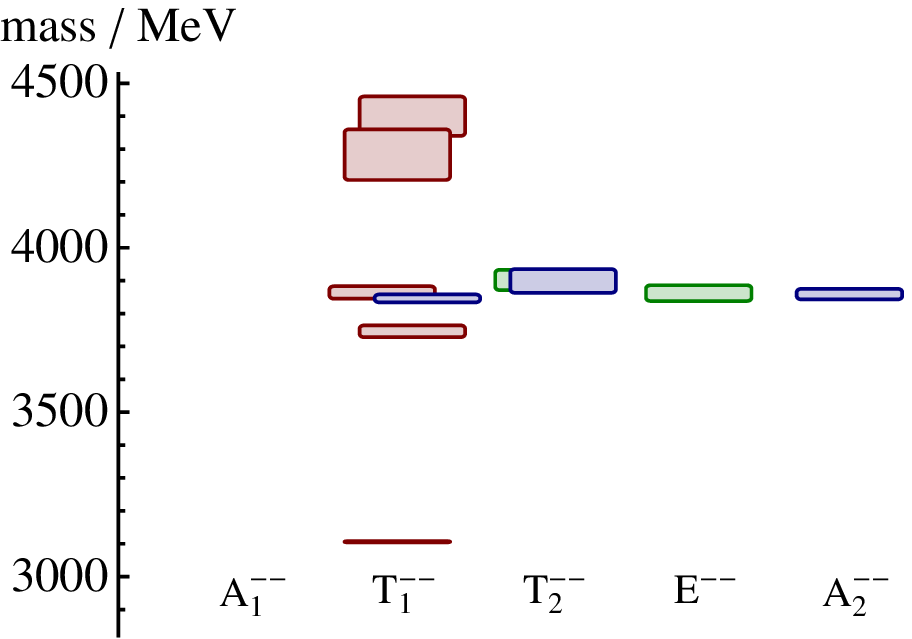}
\hfill\includegraphics[width=7.5cm]{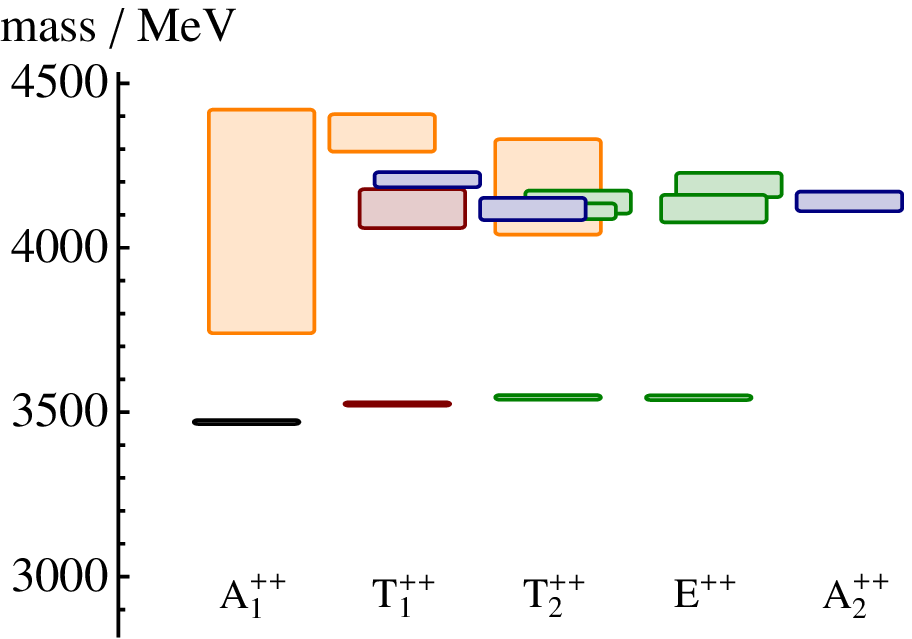}
\caption{The left- and right-hand panels show the masses calculated in
  the $J^{PC} = J^{--}~\mbox{and}~J^{++}$ channels, respectively,
  categorized according to the lattice irreps, with the color coding
  indicating their estimated continuum spin assignments where known:
  Black ($J=0$), Red ($J=1$), Green ($J=2$), Blue ($J=3$), and Orange
  (unidentified). \label{fig:charmspec}}
\end{center}
\vspace{-0.25in}
\end{figure}

As a theater for developing our methodology, we have studied
charmonium, composed of the heavier charm quark $c$ and its antiquark;
the system is particularly attractive, in that it is both
computationally far less demanding than systems composed of the light
($u,d,s$) quarks, and because there is a wealth of high-precision
experimental data.  For the study of the spectrum, performed in the
quenched approximation to QCD, we used the known continuum behavior of
our operators to enable the \textit{spins} of the states in the
different lattice irreps to be identified\cite{Dudek:2007wv},
illustrated for the $J^{PC} = J^{--}~\mbox{and}~J^{++}$ channels in
Figure~\ref{fig:charmspec}.  Lattice QCD can further enable the
electromagnetic properties of the states to be investigated.  Thus the
radiative transitions between some low-lying, non-exotic states were
studied\cite{Dudek:2006ej}, and recently this was extended to include
the radiative transition form factors for some of the higher-lying
states in the spectrum, including those with exotic quantum
numbers\cite{Dudek:2009kk}.  Most importantly, it was shown that
radiative transitions to states with exotic quantum numbers were
calculable, and furthermore the partial width for the exotic decay
$\Gamma(\eta_{c1} \rightarrow J/\psi \gamma)$ was shown to be
comparable to the non-exotic width.  The extension of this calculation
to mesons composed of light quarks, using the $N_f = 2 \oplus 1$
anisotropic clover lattices\cite{Lin:2008pr}, is in progress.

\section{Conclusions}
The calculation of the masses and properties of resonances is
essential to fully capitalize on the investment in
experimental facilities, and to provide reliable calculations to confront
the experimental results.  Thus the generation and analysis of lattices
designed for spectroscopy is a central goal of the USQCD
Collaboration's program. The calculations outlined above are but the
first stage in this program, with quark masses considerably higher
than those realized in nature.  An immediate challenge arises as we
enter the regime where resonances are unstable under the strong
interactions, and multi-hadron energies in the spectrum are expected
to be important; a novel means of efficiently computing correlation
functions in this regime has recently been
outlined\cite{Peardon:2009gh}.  
The advent of petascale computing is enabling calculations to be
performed at progressively lighter values of the light-quark masses,
and calculations with all masses fixed to their physical values are
now, finally, within reach.

\ack
This research used resources of the National Center for Computational
Sciences at Oak Ridge National Laboratory, which is supported by the
Office of Science of the Department of Energy under Contract
DE-AC05-00OR22725.  This research was supported in part by the
National Science Foundation through TeraGrid resources provided by
Pittsburgh Supercomputing Center (PSC), San Diego Supercomputing
Center (SDSC) and the Texas Advanced Computing Center (TACC), and
under the USQCD SciDAC-2 Grant.  This work used cluster resources at
Jefferson Laboratory provided under the auspicies of the USQCD
Collaboration.

MP and SR are supported by Science Foundation Ireland under research
grants 07/RFP/PHYF168 and 06/RFP/PHY061, respectively.  JB, JF and CM
are supported by National Science Foundation grant numbers
NSF-PHY-0653315 and NSF-PHY-0510020.  KJJ is supported by grant number
NSF-PHY-0704171.  Authored by Jefferson Science Associates, LLC under
US DOE Contract No.\ DE-AC05-06OR23177. The U.S. Government retains
a non-exclusive, paid-up, irrevocable, world-wide license to publish
or reproduce this manuscript for U.S. Government purposes.

\section*{References}
\bibliography{tenyear}

\end{document}